\pdfoutput=1

\documentclass[aps,prb,twocolumn,superscriptaddress]{revtex4}
\usepackage{graphicx} %graphics handler
\usepackage{dcolumn}% Align table columns on decimal point
\usepackage{bm}% bold math

\newcommand{\abgd}{ {\alpha\beta\gamma\delta} }
\newcommand{\abgdbar}{ {\bar{\alpha}\bar{\beta}\bar{\gamma}\bar{\delta}} }

\begin{document}

\title{Fermi-Edge Superfluorescence from a Quantum-Degenerate Electron-Hole Gas}
 
\author{J.-H.~Kim}
\thanks{J.-H.~Kim and G.~T.~Noe II equally contributed to this work.}
\author{G.~T.~Noe II}
\thanks{J.-H.~Kim and G.~T.~Noe II equally contributed to this work.}
\affiliation{Department of Electrical and Computer Engineering, Rice University, Houston, TX 77005, USA}

\author{S. A. McGill}
\affiliation{National High Magnetic Field Laboratory, Florida State University, Tallahassee, FL 32310, USA}

\author{Y. Wang}
\author{A. K. W\'{o}jcik}
\author{A. A. Belyanin}
\affiliation{Department of Physics and Astronomy, Texas A\&M University, College Station, TX 77843, USA}

\author{J.~Kono}
\email[]{kono@rice.edu}
\thanks{corresponding author.}
\affiliation{Department of Electrical and Computer Engineering, Rice University, Houston, TX 77005, USA}
\affiliation{Department of Physics and Astronomy, Rice University, Houston, TX 77005, USA}
\affiliation{Department of Materials Science and NanoEngineering, Rice University, Houston, TX 77005, USA}

\date{\today}

\begin{abstract}
%Nonequilibrium can be a source of order.  This rather counterintuitive statement has been proven to be true through a variety of fluctuation-driven, self-organization behaviors exhibited by out-of-equilibrium, many-body systems in nature (physical, chemical, and biological), resulting in the spontaneous appearance of macroscopic coherence.  Here, 
We report on the observation of spontaneous bursts of coherent radiation from a quantum-degenerate gas of nonequilibrium electron-hole pairs in semiconductor quantum wells.  Unlike typical spontaneous emission from semiconductors, which occurs at the band edge, the observed emission occurs at the quasi-Fermi edge of the carrier distribution.  As the carriers are consumed by recombination, the quasi-Fermi energy goes down toward the band edge, and we observe a continuously red-shifting streak.  We interpret this emission as cooperative spontaneous recombination of electron-hole pairs, or superfluorescence, which is enhanced by Coulomb interactions near the Fermi edge.  This novel many-body enhancement allows the magnitude of the spontaneously developed macroscopic polarization to exceed the maximum value for ordinary superfluorescence, making electron-hole superfluorescence even more ``super'' than atomic superfluorescence.
\end{abstract}

%\keywords{}

\maketitle


\section{Introduction}

%Currently, there is much synergism between the traditional disciplines of condensed matter physics and quantum optics.  
Recent advances in optical studies of condensed matter systems have led to the emergence of a variety of phenomena that have conventionally been studied in the realm of quantum optics, including the Rabi flopping behavior,\cite{KamadaetAl01PRL,ChoietAl10NP} the Autler-Townes splitting and dressed states,\cite{ShimanoGonokami94PRL,MulleretAl08PRL,WagneretAl10PRL} electromagnetically induced transparency,\cite{PhillipsWang02PRL} and the Mollow triplet.\cite{VamivakasetAl09NP,FlaggetAl09NP} 
%These pioneering studies have not only deepened our understanding of light-matter interactions, in general, but also introduced aspects of many-body correlations inherent in optical processes in condensed matter systems\cite{YamamotoImamoglu99Book,KiraKoch12Book}.  
%These pioneering studies have not only deepened our understanding of light-matter interactions, but also introduced aspects of many-body correlations inherent in optical processes in condensed matter systems\cite{YamamotoImamoglu99Book,KiraKoch12Book}.  
These studies have not only deepened our understanding of light-matter interactions but also introduced  aspects of {\em many-body correlations inherent in optical processes in condensed matter systems}.\cite{Toyozawa03Book,KiraKoch12Book}
%The concept of superradiance (SR)\cite{Dicke54PR,AndreevetAl80SPU,GrossHaroche82PR,ZheleznyakovetAl89SPU,ScullySvidzinsky09Science} --- cooperative radiative decay of excited dipoles --- has recently been applied to a wide range of physical situations, including Bose-Einstein condensates\cite{InouyeetAl99Science}, quantum dots\cite{ScheibneretAL07NP}, plasmonics\cite{SonnefraudetAl10ACSNano,Martin-CanoetAl10NL}, cold atoms\cite{BaumannetAl10Nature}, and lasers\cite{BohnetetAl12Nature}.  
%Superradiance (SR)\cite{Dicke54PR,AndreevetAl80SPU,GrossHaroche82PR} --- the cooperative radiative decay of excited dipoles --- has recently been found in a wide range of physical systems, including Bose-Einstein condensates\cite{InouyeetAl99Science}, quantum dots\cite{ScheibneretAL07NP}, plasmonics\cite{SonnefraudetAl10ACSNano,Martin-CanoetAl10NL}, cold atoms\cite{BaumannetAl10Nature}, and lasers\cite{BohnetetAl12Nature}.   What is common in these diverse systems is the existence of interactions among the constituent dipoles via photon exchange that determines the dynamics of the whole system.  In superfluorescence (SF)\cite{BonifacioLugiato75PRA,VrehenGibbs82Book,NoeetAl12NP}, an extreme form of SR, a giant dipole grows spontaneously, as atomic or excitonic dipoles exchange photons, and then decays superradiantly, producing a burst of coherent radiation.  

Here, we study nonequilibrium dynamics of high-density electron-hole (e-h) pairs in photo-excited semiconductor quantum wells at low temperature.  
The e-h pairs are incoherently prepared, but a macroscopic polarization spontaneously emerges and cooperatively decays, emitting a giant pulse of coherent light.  
%This phenomenon, well known as SF\cite{BonifacioLugiato75PRA,VrehenGibbs82Book} in quantum optics, is a nonequilibrium many-body process, in which order emerges in a self-organized manner via quantum fluctuations\cite{NicolisPrigogine77Book}.  
This phenomenon, known as superfluorescence (SF)\cite{BonifacioLugiato75PRA,VrehenGibbs82Book} in quantum optics, is a nonequilibrium many-body process, in which order emerges in a self-organized manner via quantum fluctuations.\cite{NicolisPrigogine77Book}
A giant dipole grows as inverted atomic dipoles interact with each other by exchanging spontaneously emitted photons.   
%As discussed in Dicke's seminal paper in 1954\cite{Dicke54PR}, as well as in the first experimental demonstrations of SF in atomic gases\cite{SkribanowitzetAl73PRL,GibbsetAl77PRL}, the resultant macroscopic polarization produced by $N$ atomic dipoles with an individual decay time of $T_1$ can cooperatively decay with an accelerated decay rate of  $NT^{-1}_1$ and an intensity $\propto N^2$.
As predicted by Dicke in 1954~\cite{Dicke54PR} and verified experimentally in atomic gases,\cite{SkribanowitzetAl73PRL,GibbsetAl77PRL} the resultant macroscopic polarization produced by $N$ atomic dipoles with an individual decay rate of $\gamma$ can cooperatively decay at an accelerated rate $N\gamma$ and an intensity $\propto N^2$.\cite{AndreevetAl80SPU,GrossHaroche82PR,ZheleznyakovetAl89SPU}

%%%%% Fig. 1 %%%%%
\begin{figure*}
\includegraphics[scale = 0.65]{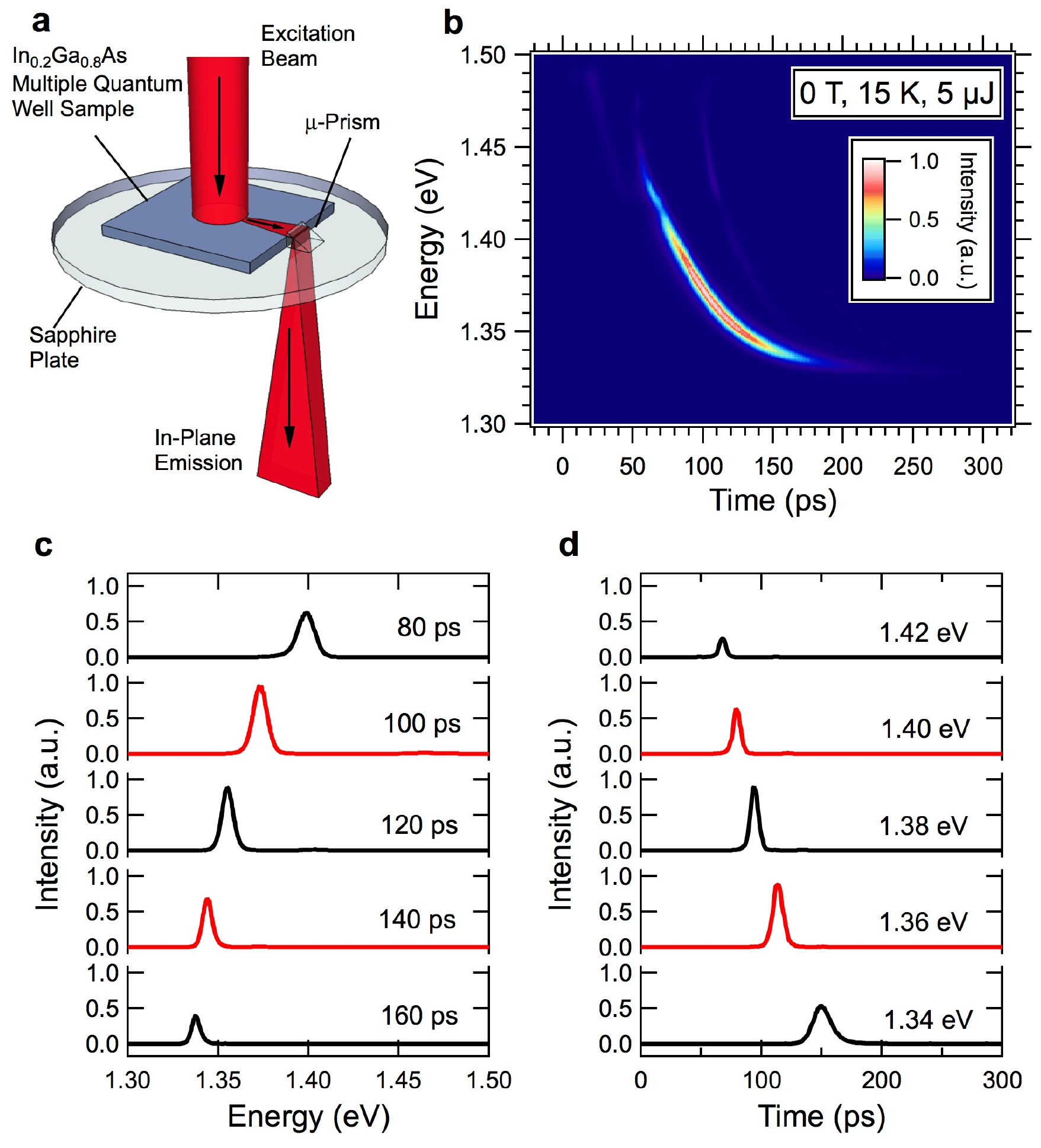}
\caption{{\bf Observation of intense ultrashort pulses of radiation from a photo-excited InGaAs quantum well sample with photon energy and time delay continuously changing with time.}  
{\bf a},~The experimental geometry. 
The in-plane emission is redirected with a micro-prism towards the collection optics.  
The sample was kept at 15~K and 0~T.  
The excitation photon energy, pulse width, and pulse energy were $\sim$1.6~eV, $\sim$150~fs, and 5~$\mu$J, respectively.  
{\bf b},~Photoluminescence intensity as a simultaneous function of time delay and photon energy.  
The peak emission red-shifts as a function of time.  
{\bf c},~Spectral slices of the map in {\bf b} for various time delays.  
{\bf d},~Temporal slices of the map in {\bf b} for various photon energies, showing pulses of radiation whose delay time with respect to the pump pulse becomes longer with decreasing photon energy.}
\end{figure*}

We demonstrate that Coulomb interactions, i.e., virtual-photon exchange, among photo-excited carriers have a profound influence on the collective superradiant decay of the dense e-h plasma.  
Contrary to a typical interband emission spectrum of semiconductors, which is concentrated near the band gap, the observed SF spectra of this Coulomb-correlated ensemble of e-h pairs show that the dominant emission originates from the recombination of electrons and holes at their respective quasi-Fermi energies.  
Consequently, we observe a red-shifting streak of SF at zero magnetic field and sequential SF bursts from different Landau levels in a quantizing magnetic field.  
The photon energy of the emitted SF mirrors the instantaneous location of the quasi-Fermi energy, which continuously decreases with time toward the band edge as the e-h pairs at the Fermi edge are consumed by SF; this dynamic red-shift is opposite to what we expect from band-gap renormalization, which should decrease as the carriers are consumed, leading to a dynamic blue-shift.  
%Overall, the many-body effects in this system are not just small corrections that require exotic conditions to be observed; rather, they are completely dominating the electron dynamics and emission spectra.  
Overall, the many-body effects in this system are not just small corrections that require exotic conditions to be observed; rather, they completely dominate the electron dynamics and emission spectra.  
%Thus, ultrabright SF, or super SF, from a dense e-h plasma is perhaps the most vivid and spectacular display of many-body physics in semiconductors.
Thus, ultrabright SF from a dense e-h plasma is one of the most vivid displays of many-body physics in semiconductors.

\section{methods}
%Put methods in here.  If you are going to subsection it, use \verb|\subsection| commands.  Methods section should be less than 800 words and if it is less than 200 words, it can be incorporated into the main text.

%\subsection{Experimental methods}

The sample studied was an undoped multiple quantum well structure, consisting of fifteen layers of 8-nm In$_{0.2}$Ga$_{0.8}$As wells and 15-nm GaAs barriers.  
By using an amplified Ti:sapphire laser with with a pulse width of $\sim$150~fs, a repetition rate of 1~kHz, and a photon energy of $\sim$1.6~eV, we generated carriers with energies higher than the band gap of the GaAs barriers.\cite{NoeetAl12NP} 
The experimental data shown in Figs.~1 and 2 was taken utilizing the optical Kerr gate method at Rice University using a 1~kHz amplified Ti:sapphire laser (Clark-MXR: CPA-2001).  
The Kerr medium used was Toluene.  
The photoluminescence was collected and imaged with off-axis parabolic mirrors onto the Kerr medium.  
A split-off portion of the excitation beam was used as the optical Kerr gate pulse.  
The time-resolved photoluminescence was measured by a CCD camera attached to a grating spectrometer after incrementally changing the time delay between the excitation and gate pulses using a one-dimensional linear stage.
The experimental data shown in Fig.~3 was taken at the National High Magnetic Field Laboratory in Tallahassee, utilizing a 17.5-T superconducting magnet.  
The sample was mounted in the Faraday geometry, where the magnetic field was parallel to the optical excitation and perpendicular to the plane of the quantum wells.  
We observed time-resolved photoluminescence using a streak camera with 2~ps time resolution. 

%%%%% Fig. 2 %%%%%
\begin{figure*}
\includegraphics[scale = 0.75]{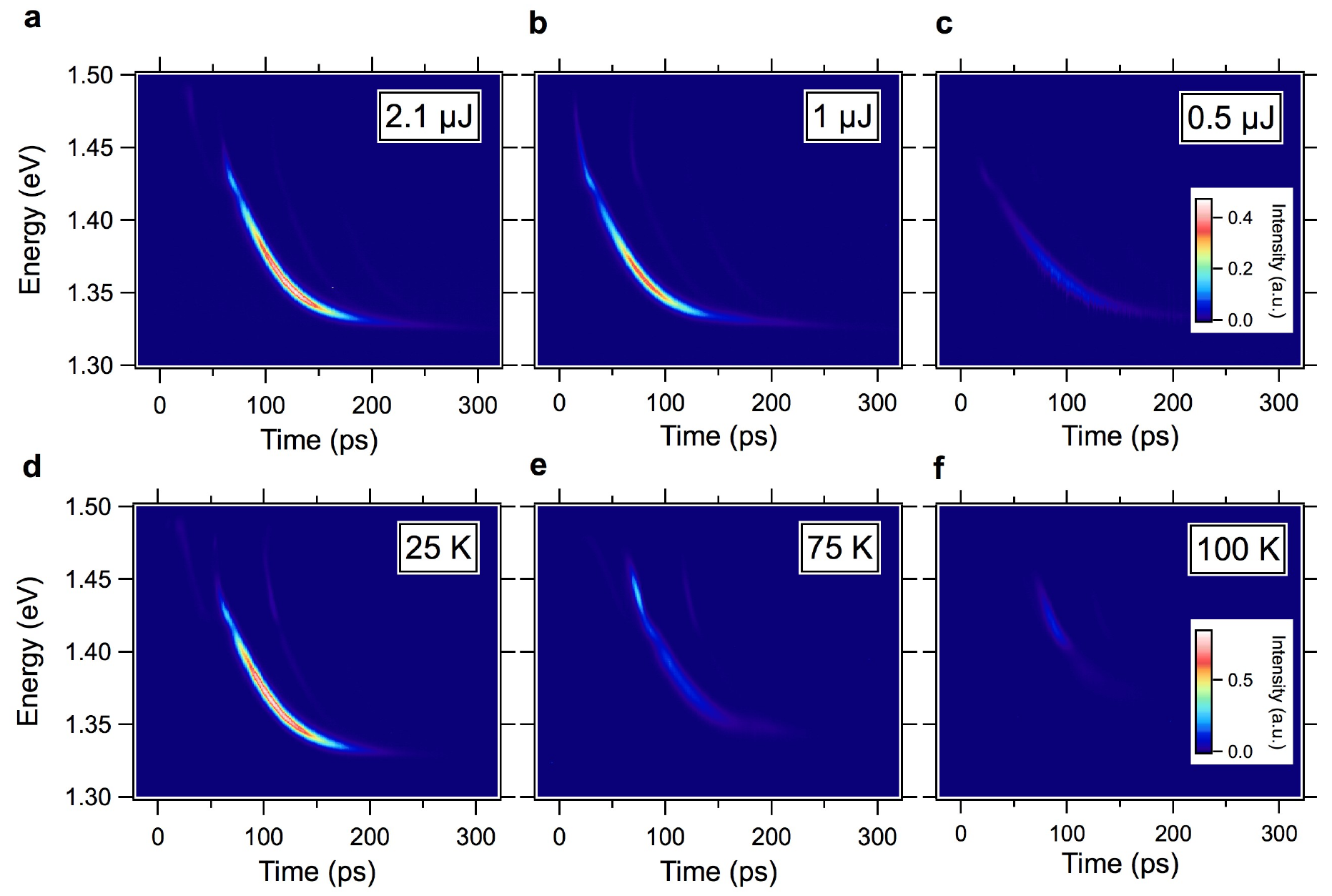}
\caption{{\bf Excitation pulse energy and temperature dependence of the observed pulsed radiation at zero magnetic field.}  
Photoluminescence intensity versus time delay and photon energy for excitation pulse energies of {\bf a},~2.1~$\mu$J, {\bf b},~1~$\mu$J, and {\bf c},~0.5~$\mu$J  at 15~K and 0~T.  
Photoluminescence intensity versus time delay and photon energy at {\bf d},~25~K, {\bf e}~75~K, and {\bf f} 100~K, with 5~$\mu$J excitation pulse energy at 0~T.  The intense pulsed emission of radiation becomes weaker with decreasing (increasing) excitation power (temperature) and eventually disappears when the excitation power (temperature) becomes too low (high).}
\end{figure*}

%%%%% Fig. 3 %%%%%
\begin{figure*}
\includegraphics[scale = 1.25]{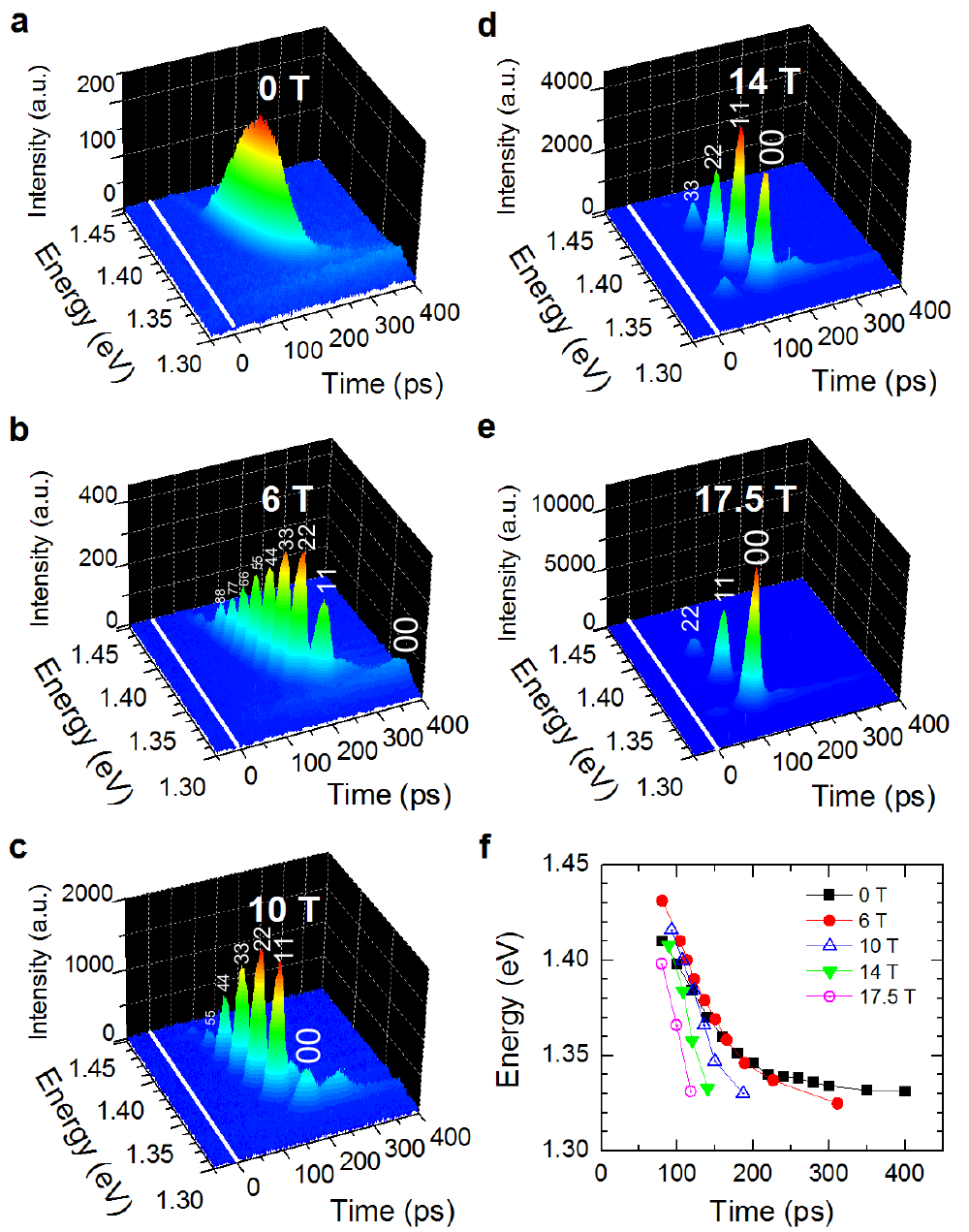}
\caption{{\bf Magnetic-field evolution of the observed pulsed coherent emission as a function of photon energy and time delay.}  
Time-resolved emission spectra at {\bf a},~0~T, {\bf b},~6~T, {\bf c},~10~T, {\bf d},~14~T, and {\bf e},~17.5~T with 2~$\mu$J of excitation pulse energy.  
Each ($N,N$) recombination is observed as a delayed burst of superfluorescence ($N$: Landau level index).
With increasing magnetic field, the number of peaks decreases, and the energy separation between adjacent peaks increases due to increasing Landau quantization energy.    
At a fixed magnetic field, the delay is longer for smaller $N$. 
Note that the $N$ = 0 state is the last to burst. 
 {\bf f}, Peak shift of emission as a function of time at different magnetic fields.}
\end{figure*}

\section{Results}

Figure 1a shows the experimental geometry used in this work.  
Photoluminescence (PL) travels in all directions, but some of the emission travels in the plane of the quantum wells, which is reflected by the micro-prism towards our collection optics.   
Figure 1b shows the result of time-resolved measurements of in-plane-emitted PL taken at 15~K at zero magnetic field with a pump pulse energy of 5~$\mu$J. 
%The dominant feature is a line of emission starting from $\sim$1.45~eV and ending at $\sim$1.325~eV, i.e., the photon energy of the emitted light is continuously changing with time.  
The dominant feature is a line of emission starting from $\sim$1.45~eV and ending at $\sim$1.325~eV, i.e., the emitted photon energy changes continuously with time.  
%There is a kink in the line at $\sim$1.42~eV close to the $E_1L_1$ transition, and the curvature of the line changes slightly at that kink.  
There is a kink in the line at $\sim$1.42~eV, which corresponds to the $E_1L_1$ transition; the curvature of the line also changes slightly at that kink.  
%\textcolor{red}{ $\leftarrow$ What was the point of this sentence?}
%Figure 1c shows some ``vertical'' slices of the data in Fig.~1a at various time delays.  
Figure 1c shows some ``vertical'' slices of the data in Fig.~1b at various time delays.  
We see that for a given time delay there is an emission peak with a spectral width of 5-10~meV, which dynamically shifts to lower energy as time passes.  
Figure 1d shows some ``horizontal'' slices of the data in Fig.~1b at various photon energies, demonstrating an ultrashort pulse of light emitted at a given photon energy at a certain time delay after excitation. 

We found that the spectral and temporal behavior of the emission line sensitively depends on the excitation pulse energy and temperature.
Figures 2a-c show time-resolved PL maps taken with different excitation pulse energies at zero magnetic field.  
%The map at 2.1~$\mu$J looks very similar to the map at 5~$\mu$J. 
The map constructed with 2.1~$\mu$J pulse energy looks very similar to the map constructed with 5~$\mu$J (Fig.~1b). 
When the power is further decreased, there is a non-monotonic temporal shift in the line of emission. 
For a given photon energy close to the middle of the line, say 1.37~eV, we see that the line moves to earlier time from 2.1 to 1~$\mu$J and then back to a later time at 0.5~$\mu$J excitation pulse energy with a change in curvature.  
At the highest photon energy for strong emission in the line, at $\sim$1.45 eV, the emission moves to earlier time delays with decreasing power and then stays there for the lowest power.  
%For all excitation powers, the line ends at the same energy, 1.325~eV, which corresponds to the $E_1H_1$ band-edge.
For all excitation powers, the emission line ends at 1.325~eV, which corresponds to the $E_1H_1$ band-edge. 
%We also varied the temperature while keeping the excitation pulse energy constant at 5~$\mu$J, as shown in Figs.~2d-f.  
We also varied the temperature while fixing the excitation pulse energy at 5~$\mu$J, as shown in Figs.~2d-f.  
With increasing temperature, there is a smearing of the emission line at the lowest photon energies of the line, until all of the emission from the $E_1H_1$ contribution of the line is `washed out', and only the slightest signal at the $E_1L_1$ portion remains at 100~K.  
%We must remember that the energy gap in semiconductors red-shifts with increasing temperature when trying to identify which part of the line corresponds to which subband.  
It is clear that the emission burst moves to later times as it is `washed out' at high temperatures.
%\textcolor{red}{$\leftarrow$ I do not like this paragraph. It simply describes the experimental data without explaining why these descriptions are important.}

%%%%% Fig. 4 %%%%%
\begin{figure}
\includegraphics[scale = 0.47]{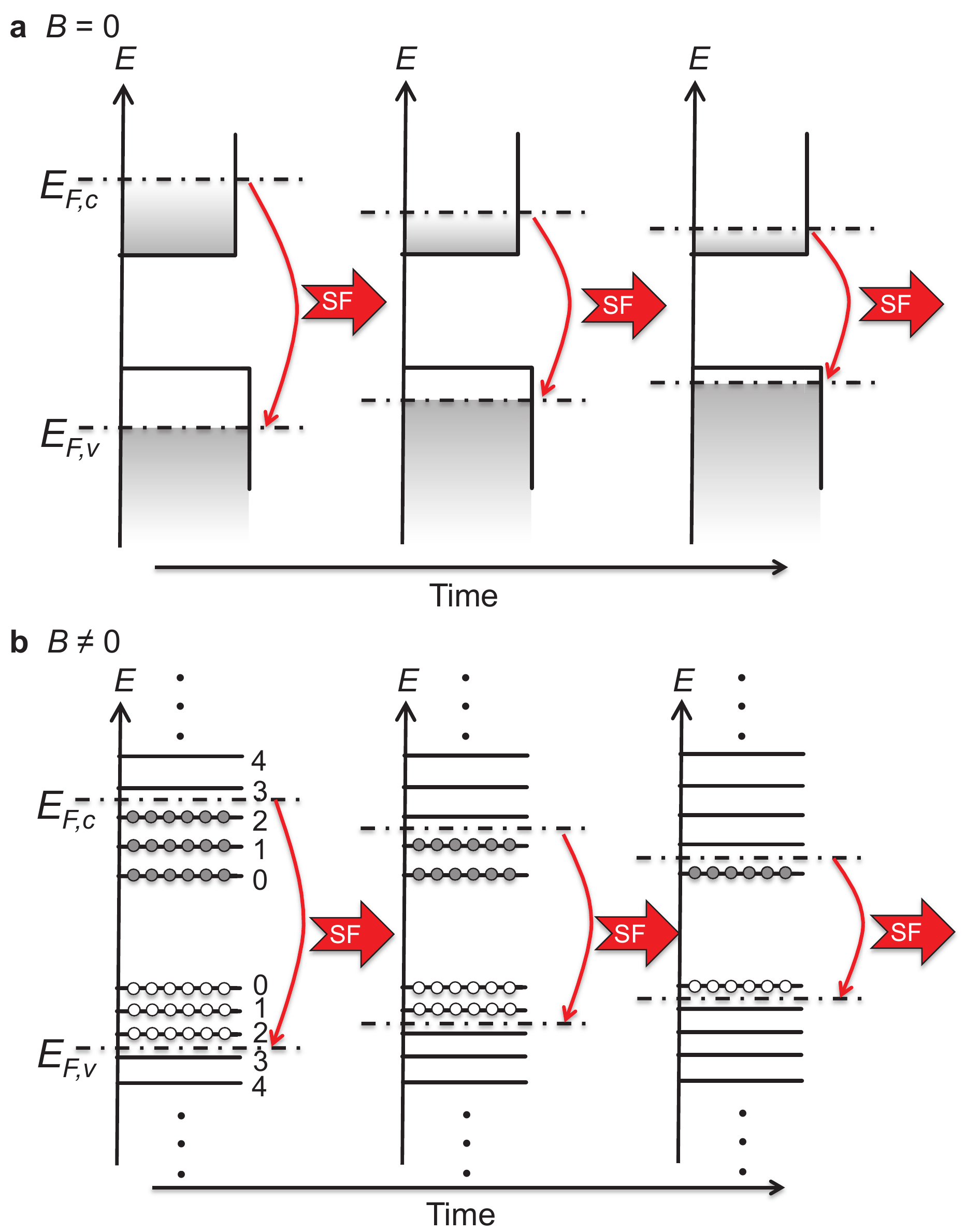}
\caption{{\bf Interpretation of the burst of emission with dynamically red-shifting wavelength: coherent Fermi-edge emission via ultrafast superradiant recombination of an electron-hole plasma}. 
 {\bf a},~Zero magnetic field.  Carriers near the instantaneous quasi-Fermi energies are consumed through ultrafast cooperative recombination, due to many-body enhancement of gain,\cite{Schmitt-RinketAl86PRB} leading to both a burst of radiation with continuously red-shifting wavelength and a continuously decreasing Fermi energy towards the band edge. 
 {\bf b},~Finite magnetic field.  
 Electron-hole pairs at the highest occupied Landau levels recombine first, again due to many-body enhancement of gain, leading to sequential bursts of superfluorescece from higher to lower Landau levels toward the (00) level.}
\end{figure}

The emission spectrum and dynamics drastically change when a magnetic field perpendicular to the quantum well plane is applied.
Figures 3a-e show streak camera images of emission as a function of photon energy and time delay at different magnetic fields.  
With increasing magnetic field, the number of peaks decreases, and the energy separation between adjacent peaks increases due to increasing Landau quantization energy (i.e., the cyclotron energy).  
Previously we demonstrated the superradiant nature of the individual emission peaks by  streak-camera and pump-probe measurements.\cite{NoeetAl12NP}
%Here we observe that at a given magnetic field the delay is longer for emission with a smaller Landau-level index $N$, and the ($NN$) = (00) SF emission occurs only after the higher-energy SF emissions occur.  
Here we observe that at a given magnetic field the delay is longer for emission from lower Landau levels, and the ($NN$) = (00) SF emission occurs only after the higher-energy SF emissions occur.  
This means that the relative timing of the bursts coming from different Landau levels is not random.  
%Rather, these data clearly indicate that e-h pairs in the highest occupied energy states near the quasi-Fermi edge at a given time always recombine first; e-h pairs in lower and lower energy states then emit bursts in a sequential manner in time.  
Rather, these data clearly indicate that e-h pairs in the highest occupied energy states near the quasi-Fermi edge at a given time always recombine first; e-h pairs in lower and lower energy states then emit bursts sequentially.  
Figure 3f summarizes the peak positions of the SF bursts as a function of photon energy and time.

\begin{figure*}
\includegraphics[scale = 0.75]{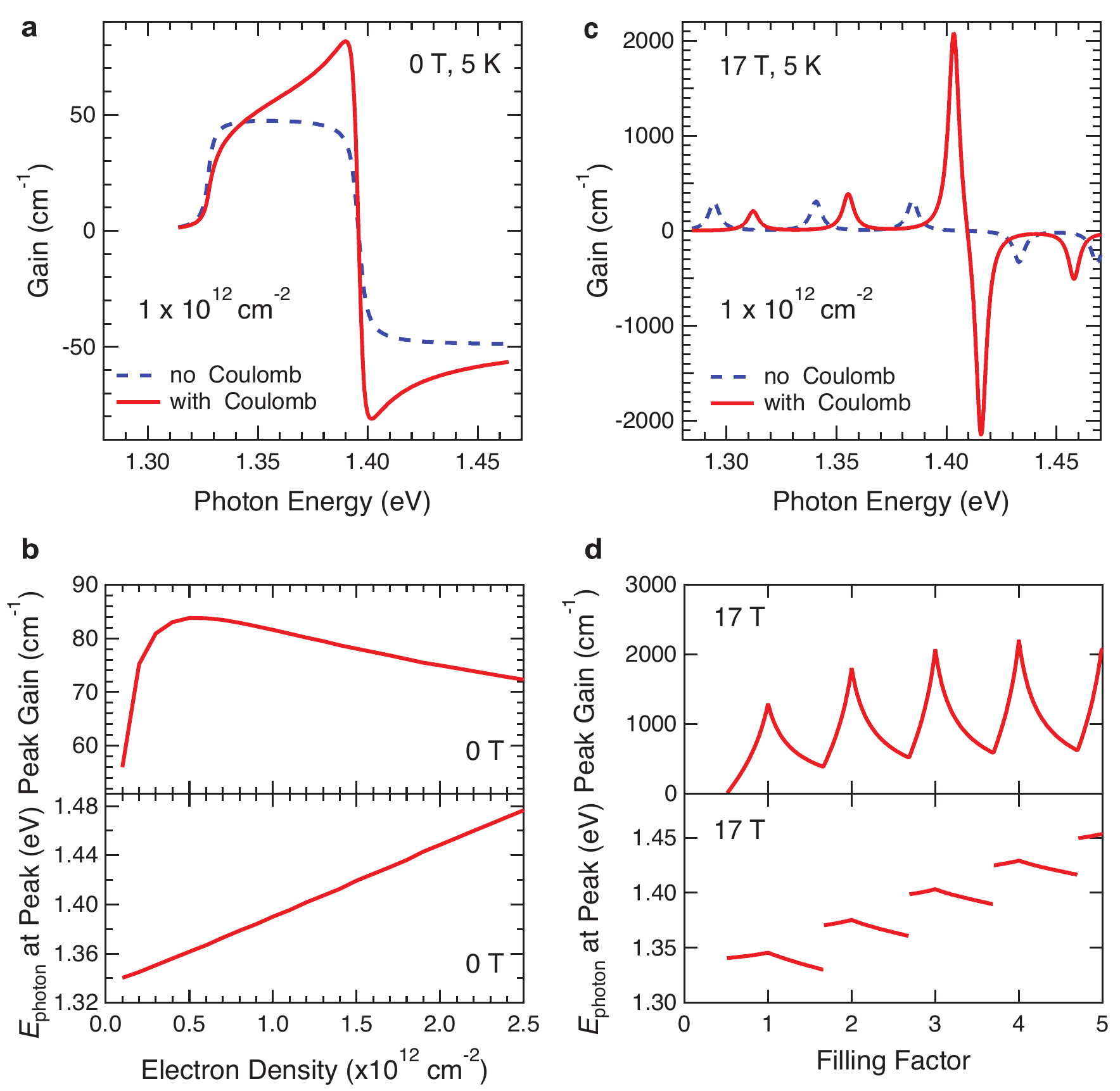}
\caption{{\bf Theoretical calculations of Coulomb-induced many-body enhancement of gain at the Fermi energy at zero magnetic field and 17~T.}  {\bf a}, Gain spectrum for the InGaAs sample without a magnetic field, calculated using Eq.~(\ref{Eq::chimain}) (solid line), in comparison with the spectrum obtained by replacing $\chi_\alpha(\omega) \rightarrow \chi^0_\alpha(\omega)$, i.e., neglecting all Coulomb effects except band-gap renormalization (dashed line).  
Separate Fermi distributions for electrons and holes of density $1 \times 10^{12}$~cm$^{-2}$ and temperature 5~K are assumed. A relaxation rate of 2~meV is assumed.  
{\bf b}, Peak gain (upper panel) and peak gain energy (lower panel) as a function of e-h density at zero magnetic field. 
Other parameters are the same as in {\bf a}.  
{\bf c},~Calculated gain spectrum in a magnetic field of 17~T (solid line), in comparison with the spectrum obtained by replacing $\chi_\alpha(\omega) \rightarrow \chi^0_\alpha(\omega)$, i.e., neglecting all Coulomb effects except band-gap renormalization (dashed line).  
A filling factor $\nu = 3$ and a temperature of 5~K are assumed. 
A relaxation rate of 3~meV is adopted.  
{\bf d},~Peak gain (upper panel) and peak gain energy (lower panel) at 17~T as a function of filling factor, defined as the number of filled Landau levels. Other parameters are the same as in {\bf c}.} 
\end{figure*}

\section{Discussion}

We interpret these phenomena in terms of Coulomb enhancement of gain near the Fermi energy in a high-density e-h system, which results in a preferential SF burst near the Fermi edge, as schematically shown in Fig.~4.  
After relaxation and thermalization, the photo-generated carriers form degenerate Fermi gases with respective quasi-Fermi energies inside the conduction and valence bands. 
The recombination gain for the e-h states just below the quasi-Fermi energies is predicted to be enhanced due to Coulomb interactions among carriers,\cite{Schmitt-RinketAl86PRB} which causes a SF burst to form at the Fermi edge most easily.  
As a burst occurs, a significant population is depleted, resulting in a decreased Fermi energy.  Thus, as time goes on, the Fermi level moves toward the band edge continuously.  
This results in a continuous line of SF emission at zero field (Figs.~1 and 2) and a series of sequential SF bursts in a magnetic field (Fig.~3).

We model the recombination dynamics of the photo-excited e-h plasma using semiconductor Bloch equations derived from a general two-band e-h Hamiltonian in the Hartree-Fock approximation (see Appendix for more details). 
At the linear stage of SF, when the field grows exponentially, the gain spectrum is given by $g(\omega) = \frac{4\pi\omega}{n_b c} \mathrm{Im} [\chi(\omega)] $,  where $n_b$ is the background refractive index, $c$ is the speed of light, $\chi(\omega) = {1 \over V} \sum_\alpha \mu_\alpha^\ast \chi_\alpha(\omega)$ is the optical susceptibility, and $V$ is the normalization volume.  
The functions $\chi_{\alpha}(\omega)$ satisfy a set of linear equations
%%%
\begin{eqnarray}
\label{Eq::chimain}
\chi_\alpha(\omega) = \chi^0_\alpha(\omega) \left[ 1 + {1 \over \mu_\alpha} \sum_{\beta} V_{\alpha\beta\beta\alpha} \chi_\beta(\omega) \right],
\end{eqnarray}
%%%
where
%%%
%\begin{eqnarray}
%\label{Eq::chi0main}
%\chi^0_\alpha(\omega) = \frac{\mu_\alpha \left( n^e_\alpha + n^h_\alpha - 1 \right)} {\hbar\omega - \left( E_g^0 + E^{eR}_\alpha + E^{hR}_\alpha \right) + i\hbar\gamma_\alpha}~, 
%\end{eqnarray}
\begin{eqnarray}
\label{Eq::chi0main}
\chi^0_\alpha(\omega) := \frac{\mu_\alpha \left( n^e_\alpha + n^h_\alpha - 1 \right)} {\hbar\omega - \left( E_g^0 + E^{eR}_\alpha + E^{hR}_\alpha \right) + i\hbar\gamma_\alpha} \, , 
\end{eqnarray}
%%%
each Greek subscript ($\alpha,\beta$, ...) denotes a set of quantum numbers for a given single-particle state (e.g., wave vector, Landau level index, and spin), $\mu_{\alpha}$ is the dipole matrix element of the optical transition between electron and hole states with index $\alpha$, $E^{eR}_\alpha$ = $\left( E^e_\alpha - \sum_\beta V_{\alpha\beta\beta\alpha} n^e_\beta \right)$ and $E^{hR}_\alpha$ = $\left( E^h_\alpha - \sum_\beta V_{\alpha\beta\beta\alpha} n^h_\beta \right)$ are the renormalized energies of single-particle states $E^{e,h}_\alpha$ , $n_\alpha^e$ and $n_\alpha^h$ are e-h occupation numbers, and $\gamma_{\alpha}$ is the phenomenological dephasing term for the interband polarization. 
Matrix elements $V_{\alpha \beta \gamma\delta}$ of the screened Coulomb interaction are specified in Appendix A; screening is calculated using the Lindhard formula. 

An example of Coulomb-induced modification of gain for quantum wells at zero magnetic field is shown in Fig.~5a (solid line), together with a gain spectrum neglecting all Coulomb effects except band gap renormalization (dashed line); the latter was obtained by replacing $\chi_\alpha(\omega)$ by $\chi^0_\alpha(\omega)$.  
It is seen that Coulomb interactions lead to an enhancement of gain just below the energy that corresponds to the difference between the quasi-Fermi levels of electrons and heavy holes.  
Previously, a related effect of ``Fermi-edge singularity'' has been observed in the spontaneous PL spectra of n-doped quantum wells in a steady state.\cite{SkolnicketAl87PRL}  
In the present case, the many-body gain enhancement is completely due to a nonequilibrium photo-excited e-h plasma.  
Stimulated emission occurs in the quantum well plane, and the light intensity {\em grows exponentially}, both in space and time.  
As a result, the rather broad many-body enhancement in the gain spectrum around Fermi energy translates into a sharp peak in the instantaneous intensity spectrum.  
The subsequent time evolution of the spectrum is dominated by an ultrafast collective recombination process: the peak continuously follows the red-shift of the quasi-Fermi level as the carriers at the Fermi edge are consumed by the SF.  
This behavior, observed in our samples according to Fig.~2, is in agreement with Fig.~5b, which shows the calculated evolution of the peak gain and peak gain energy as a function of e-h pair density.  
Furthermore, the highest gain, which leads to the fastest decay, is seen to be achieved at some intermediate density, which explains the observed non-monotonic temporal shift as a function of pump power (Figs.~2a-c).

In a strong magnetic field, the gain spectrum exhibits strong peaks when the Landau level filling factor is an integer, and for a given filling factor, the gain is largest for the highest filled Landau level (Figs.~5c and 5d).  
A snapshot of the gain for a fixed filling factor $\nu = 3$, corresponding to three filled Landau levels, is shown in Fig.~5c for a magnetic field of 17~T.  
It can be seen that the peak gain for e-h pairs at the $N$ = 3 Landau level is much higher than that for completely filled, lower Landau levels.  
%It should be also noted that the peak gain value is strongly enhanced as compared to quantum wells without a magnetic field due to an increase in the transition matrix element and density of states.  
Note that the peak gain value is strongly enhanced compared to quantum wells without a magnetic field due to an increase in the transition matrix element and density of states.  
This provides a natural explanation for the trend observed in Fig.~3f, i.e., SF develops faster in a stronger magnetic field.    
Figure 5d shows the calculated peak gain and peak gain energy as a function of filling factor at a fixed magnetic field of 17~T.  The peculiar many-body dynamics of the peak gain lead to isolated SF bursts that are fired consecutively from higher to lower Landau levels, as observed in Fig.~3.

\section{Summary}

%In summary, the results of this experiment-theory combined study not only provides new insight into the non-equilibrium dynamics of Coulomb-correlated e-h pairs in semiconductors but also opens up new possibilities of controlling, and {\em enhancing}, collective emission properties of many-body states via Coulomb interactions.  
In summary, the results of this study not only provide new insight into the nonequilibrium dynamics of Coulomb-correlated electron-hole pairs in semiconductors but also open up new possibilities of controlling, and {\em enhancing}, collective emission properties of many-body states.  
Specifically, we showed that superfluorescence, a well-known phenomenon in quantum optics of atoms based on photon exchange between inverted atomic dipoles, takes a new turn when it occurs in a condensed matter system, where Coulomb correlations (i.e., virtual-photon exchange) create enormous gain concentrated at the Fermi edge, which becomes better defined at lower temperatures.  
Thus, this work demonstrates a unique method of producing ultrashort pulses of radiation from a semiconductor, based on the existence of Fermi-degenerate, nonequilibrium electrons and holes.

%\pagebreak

\appendix
\section{Theoretical Modeling of Coulomb-Enhancement of Gain at the Fermi Edge}

We used the semiconductor Bloch equations (SBEs) to study SF from a high-density electron-hole (e-h) plasma in the presence of many-body Coulomb interactions.  The usual form of the SBEs\cite{HaugKoch04Book} is for a bulk semiconductor or a 2D electron gas, when the states can be labeled by a 3D or 2D wave vector $\vec{k}$. Here we rederive SBEs following the same basic approximations but in a more general form, which accommodates the effects of a finite well width and the quantization of motion in a strong magnetic field. 

We begin with a general Hamiltonian in the two-band approximation and e-h representation,
%%%
\begin{widetext}
\begin{eqnarray}
{\cal H} &=& \sum_\alpha{ \left[ \left(E_g^0 + E^e_\alpha \right) a_\alpha^\dagger a_\alpha + E^h_\alpha b_{\bar{\alpha}}^\dagger b_{\bar{\alpha}} \right]} \nonumber \\
&+& \frac{1}{2} \sum_\abgd \left( V^{ee}_\abgd a_\alpha^\dagger a_\beta^\dagger a_\delta a_\gamma + V^{hh}_\abgdbar b_{\bar{\alpha}}^\dagger b_{\bar{\beta}}^\dagger b_{\bar{\delta}} b_{\bar{\gamma}} + 2 V^{eh}_{\alpha\bar{\beta}\gamma\bar{\delta}} a_\alpha^\dagger b_{\bar{\beta}}^\dagger b_{\bar{\delta}} a_\gamma \right) \nonumber \\
 &-& {\cal E}(t) \sum_\alpha \left( \mu_\alpha a_\alpha^\dagger b_{\bar{\alpha}}^\dagger + \mu_\alpha^\ast b_{\bar{\alpha}} a_\alpha \right) ~,
\end{eqnarray}
\end{widetext}
%%%
where $E_g^0$ is the unperturbed bandgap, $a_\alpha^\dagger$ and $b_{\bar{\alpha}}^\dagger$ are the creation operators for the electron state $\alpha$ and hole state $\bar{\alpha}$, respectively, ${\cal E}(t)$ is the optical field, $\mu_\alpha$ is the dipole matrix element, and $V_\abgd$ are Coulomb matrix elements, for example, $V^{ee}_\abgd$ = $\int d \vec{r}_1 \int d\vec{r}_2 \Psi^{e\ast}_\alpha(\vec{r}_1) \Psi^{e\ast}_\beta(\vec{r}_2) \frac{e^2}{\epsilon |\vec{r}_1 - \vec{r}_2|} \Psi^e_\gamma(\vec{r}_1) \Psi^e_\delta(\vec{r}_2)$. Here we denote the hole state which can be recombined with a given electron state $\alpha$ optically by $\bar{\alpha}$, and assume that there is a one-to-one correspondence between them. For the interband Coulomb interaction, $V^{eh}_{\alpha\bar{\beta}\gamma\bar{\delta}} a_\alpha^\dagger b_{\bar{\beta}}^\dagger b_{\bar{\delta}} a_\gamma$ is the only non-zero matrix element due to the orthogonality between the Bloch functions of the conduction and valence bands.\cite{VaskoKuznetsov99Book}  The electron and hole wave functions can be written as $\Psi^e_\alpha(\vec{r})$ = $\psi^e_\alpha(\vec{r}) u_{c0}(\vec{r})$ and $\Psi^h_{\bar{\alpha}}(\vec{r})$ = $\psi^h_{\bar{\alpha}}(\vec{r}) u^\ast_{v0}(\vec{r})$, respectively. In the problems we study, the conduction band and valence band states connected by an optical transition always have the same envelope wave function, so we take $\psi^h_{\bar{\alpha}}(\vec{r})$ = $\psi^{e\ast}_\alpha(\vec{r})$. Then the Coulomb matrix elements are related with each other through $V^{hh}_\abgdbar = V^{ee}_{\gamma\delta\alpha\beta}$ and $V^{eh}_{\alpha\bar{\beta}\gamma\bar{\delta}} = - V^{ee}_{\alpha\delta\gamma\beta}$, and we can drop the superscript by defining $V_{\alpha\beta\gamma\delta}$ $\equiv$ $V^{ee}_{\alpha\beta\gamma\delta}$. 

Using the above Hamiltonian, we can obtain the equations of motion for the distribution functions $n_\alpha^e$ = $\langle a_\alpha^\dagger a_\alpha \rangle$ and $n_\alpha^h$ = $\langle b_{\bar{\alpha}}^\dagger b_{\bar{\alpha}} \rangle$, and the polarization $P_\alpha$ = $\langle b_{\bar{\alpha}} a_\alpha \rangle$. Using the Hartree-Fock approximation (HFA) and the random phase approximation (RPA), we arrive at the SBEs:
%%%
\begin{widetext}
\begin{eqnarray}
\label{Eq::SBEs}
i\hbar {d\over d t} P_\alpha &=& \left( E_g^0 + E^{eR}_\alpha + E^{hR}_\alpha \right) P_\alpha + \left( n^e_\alpha + n^h_\alpha - 1 \right) \left[ \mu_\alpha {\cal E}(t) + \sum_\beta V_{\alpha\beta\beta\alpha} P_\beta \right] + \left. i\hbar {d\over d t} P_\alpha \right|_{\rm scatt} ~, \\
 \hbar {d\over d t} n^e_\alpha &=& - 2 ~\mathrm{Im} \left[ \left( \mu_\alpha {\cal E}(t) + \sum_\beta V_{\alpha\beta\beta\alpha} P_\beta \right) P_\alpha^\ast \right] + \left. \hbar {d\over d t} n^e_\alpha \right|_{\rm scatt} ~, \\
 \hbar {d\over d t} n^h_\alpha &=& - 2 ~\mathrm{Im} \left[ \left( \mu_\alpha {\cal E}(t) + \sum_\beta V_{\alpha\beta\beta\alpha} P_\beta \right) P_\alpha^\ast \right] + \left. \hbar {d\over d t} n^h_\alpha \right|_{\rm scatt} ~,
 \end{eqnarray}
 \end{widetext}
 %%%
where $E^{eR}_\alpha$ = $\left( E^e_\alpha - \sum_\beta V_{\alpha\beta\beta\alpha} n^e_\beta \right)$ and $E^{hR}_\alpha$ = $\left( E^h_\alpha - \sum_\beta V_{\alpha\beta\beta\alpha} n^h_\beta \right)$ are the renormalized energies, and the scattering terms account for higher-order contributions beyond the HFA and other scattering processes such as scattering with LO-phonons. 

These equations, together with Maxwell's equations for the electromagnetic field, can be applied to study the full nonlinear dynamics of interaction between the e-h plasma and radiation. Here we derive the gain for given carrier distributions $n_\alpha^e$ and $n_\alpha^h$, which was used to plot Fig.~5. Assuming a monochromatic and sinusoidal time dependence for the field ${\cal E}(t)$ = ${\cal E}_0 e^{-i\omega t}$ and the polarization $P_\alpha(t)$ = $P_{0\alpha} e^{-i\omega t}$, we can find $P_\alpha$ from Eq.~(\ref{Eq::SBEs}) and define the quantity $\chi_\alpha(\omega)$ = $P_{0\alpha}/{\cal E}_0$, which satisfies the equation below:
%%%
\begin{eqnarray}
\label{Eq::chi}
\chi_\alpha(\omega) = \chi^0_\alpha(\omega) \left[ 1 + {1 \over \mu_\alpha} \sum_{\beta} V_{\alpha\beta\beta\alpha} \chi_\beta(\omega) \right] ~,
\end{eqnarray}
%%%
where
%%%
\begin{eqnarray}
\label{Eq::chi0}
\chi^0_\alpha(\omega) = \frac{\mu_\alpha \left( n^e_\alpha + n^h_\alpha - 1 \right)} {\hbar\omega - \left( E_g^0 + E^{eR}_\alpha + E^{hR}_\alpha \right) + i\hbar\gamma_\alpha} ~.
\end{eqnarray}
%%%
Here we have written the dephasing term phenomenologically as $d P_\alpha / d t |_{\rm scatt} = - \gamma_\alpha P_\alpha$. The optical susceptibility is then
%%%
\begin{eqnarray}
\label{Eq::chiw}
\chi(\omega) = {1 \over V} \sum_\alpha \mu_\alpha^\ast \chi_\alpha(\omega) ~,
\end{eqnarray}
%%%
where $V$ is the normalization volume. The gain spectrum is given by\cite{HaugKoch04Book}
%%%
\begin{eqnarray}
\label{Eq::gainw}
g(\omega) = \frac{4\pi\omega}{n_b c} \mathrm{Im} [\chi(\omega)] ~,
\end{eqnarray}
%%%
where $n_b$ is the background refractive index, and $c$ is the speed of light.  We use the above general results to analyze optical properties under different conditions.

In a quantum well of thickness  $L_{\rm w}$, the envelope functions for  electrons and holes are  $\psi^{e,h}_{n,\vec{k}}(\vec{r})$ = $\varphi_n(z)\exp \left(i \vec{k} \cdot \vec{\rho} \right)/\sqrt{A}$, where $\vec{\rho}$ = $(x,y)$, $\varphi_n(z)$ is the envelope wave function in the growth direction for the $n$-th subband, and $A$ is the normalization area.  To calculate the Coulomb matrix element $V_{\alpha\beta\beta\alpha}$, we define $\tilde{V}_{\alpha\beta}$ $\equiv$ $V_{\alpha\beta\beta\alpha}$ and put $\alpha$ = $\left\{ n, \vec{k}, s \right\}$, $\beta$ = $\left\{ n', \vec{k}', s' \right\}$, where $s$ denotes the spin quantum index. Then one gets
%%%
\begin{eqnarray}
\tilde{V}_{n,\vec{k},s;n',\vec{k}',s'} = V^{2D}(q) F_{n n' n' n}(q) \delta_{s s'} ~,
\end{eqnarray}
%%%
where $q$ = $|\vec{q}|$ = $|\vec{k}-\vec{k}'|$, $V^{2D}(q)$ = $2\pi e^2 / \epsilon A q$, $\epsilon$ is the dielectric function, and the form factor $F_{n n' n' n}(q)$ is defined as
%%%
\begin{widetext}
\begin{eqnarray}
\label{Eq::Formfactor}
F_{n1, n2, n3, n4}(q) = \int d z_1 \int d z_2 \varphi_{n1}^\ast(z_1) \varphi_{n2}^\ast(z_2) \exp \left( - q \left| z_1 - z_2 \right| \right) \varphi_{n3}(z_1) \varphi_{n4}(z_2) ~.
\end{eqnarray}
\end{widetext}
%%%
Throughout the paper, we assume that only the lowest subband for electrons and holes is occupied. In this case, we can define $\tilde{V}(q)$ = $V^{2D}(q) F_{1111}(q)$. The dielectric function $\epsilon(\vec{q},\omega)$, which describes the screening of the Coulomb potential, is given by the Lindhard formula for a pure 2D case;\cite{HaugKoch04Book} it can be generalized to the quasi-2D case as 
\begin{eqnarray}
\label{Eq::Lindhard}
\epsilon(\vec{q},\omega) = 1 + \tilde{V}(q) \left( \Pi_e(\vec{q},\omega) + \Pi_h(\vec{q},\omega) \right) ~,
\end{eqnarray}
%%%
where $\Pi_{e(h)}(\vec{q},\omega)$ is the polarization function of an electron or hole, which is given by
%%%
\begin{eqnarray}
\Pi(\vec{q},\omega) = 2 \sum_{\vec{k}} \frac{n_{\vec{k}+\vec{q}} - n_{\vec{k}}}{\omega + i 0^+ - E_{\vec{k}+\vec{q}} + E_{\vec{k}}} ~.
\end{eqnarray}
%%%
Here, we dropped the subscripts $e$ or $h$, $n_{\vec{k}}$ is  the distribution function, the factor of 2 accounts for the summation over spin, and the spin index is suppressed. For simplicity, we will choose the static limit, namely, $\omega$ = 0. 

Given the dielectric function $\epsilon(q,0)$, the screened Coulomb matrix element is $\tilde{V}_s(q)$ = $\tilde{V}(q)/\epsilon(q,0)$. For simplicity, we will still write it as $\tilde{V}(q)$. Applying Eq. (\ref{Eq::chi}) to the case above, we get the equation for $\chi_{\vec{k}}(\omega)$: 
%%%
\begin{eqnarray}
\label{Eq::chiQW}
\chi_{\vec{k}}(\omega) = \chi^0_{\vec{k}}(\omega) \left[ 1 + {1 \over \mu_{\vec{k}}} \sum_{\vec{k}'} \tilde{V}\left( \left| \vec{k}-\vec{k}' \right| \right) \chi_{\vec{k}'}(\omega) \right]
\end{eqnarray}
%%%
where $\chi^0_{\vec{k}}(\omega)$ becomes
%%%
\begin{eqnarray}
\chi^0_{\vec{k}}(\omega) = \frac{\mu_{\vec{k}} \left( n^e_{\vec{k}} + n^h_{\vec{k}} - 1 \right)} {\hbar\omega - \left( E_g^0 + E^{eR}_{\vec{k}} + E^{hR}_{\vec{k}} \right) + i\hbar\gamma_{\vec{k}}} ~.
\end{eqnarray}

To solve Eq. (\ref{Eq::chiQW}), we notice that $\chi^0_{\vec{k}}(\omega)$ does not depend on the direction of $\vec{k}$, so $\chi_{\vec{k}}(\omega)$ will not depend on it, either. Then, after converting the summation in Eq.~(\ref{Eq::chiQW}) into the integral, the integration over the azimuthal angle is acting on $\tilde{V}\left( \left| \vec{k}-\vec{k}' \right| \right)$ only. If we define
%%%
\begin{widetext}
\begin{eqnarray}
 \tilde{V}\left(k,k' \right) = {1\over 2\pi} \int_0^{2\pi} d \phi \tilde{V}\left( \sqrt{k^2+k^{'2} - 2 k k' \cos \phi} \right) ~,
\end{eqnarray}
%%%
then Eq.~(\ref{Eq::chiQW}) can be written as
%%%
\begin{eqnarray}
\chi_k(\omega) = \chi^0_k(\omega) \left[ 1 + {A \over 2 \pi \mu_k} \int_0^\infty k' d k'  \tilde{V}\left( k, k' \right) \chi_k'(\omega) \right] ~.
\end{eqnarray} 
\end{widetext}
%%%
After discretizing the integral, we have a system of linear equations for $\chi_k(\omega)$, which can be solved by using LAPACK.\cite{LAPACK-guide} The band structure for our sample consisting of undoped 8-nm In$_{0.2}$Ga$_{0.8}$As wells and 15-nm GaAs barriers on a GaAs substrate is calculated using the parameters given by Vurgaftman {\it et al.}\cite{VurgaftmanetAl01JAP} The strain effect is included using the results of Sugawara {\it et al.}\cite{SugawaraetAl93PRB} Examples of calculated gain spectra are shown in Figs.~5a and 5b.

For a quantum well structure in a strong perpendicular magnetic field, the electronic states are fully quantized. Considering only the lowest subband in the quantum well, the equation for the susceptibility is written as
%%%
\begin{eqnarray}
\chi_{\nu,s} = \chi^0_{\nu,s} \left[ 1 + {1 \over \mu_{\nu,s}} \sum_{\nu'} V_{\nu,\nu'} \chi_{\nu',s}\right] ~,
\end{eqnarray}
%%%
where $\nu$ is the Landau level index, $s$ is the spin index, and $V_{\nu,\nu'}$ is the Coulomb matrix element given by
%%%
\begin{widetext}
\begin{eqnarray}
V_{\nu,\nu'} = \frac{e^2}{2\pi\epsilon} \int_0^{2\pi} d\theta \int_0^{\infty} d q \left| \int d x e^{i q x \cos\theta} \phi_{\nu} (x) \phi_{\nu'}^\ast(x + q a_H^2 \sin\theta) \right|^2 ~,
\end{eqnarray}
\end{widetext}
where $\phi_{\nu}(x)$ is the $x$-dependent part of the wavefunction of the $\nu$-th Landau level and $a_H^2$ = $\hbar c/e B$. The renormalized electronic energies in the expression for $\chi^0_{\nu,s}$ are
%%%
\begin{eqnarray}
E_{\nu,s}^{eR} = E_{\nu,s}^e - \sum_{\nu'} V_{\nu,\nu'} n_{\nu'}^e ~,
\end{eqnarray} 
%%%
and a similar equation holds for holes. The gain is calculated as
%%%
\begin{eqnarray}
\label{gainmag}
g(\omega) = \frac{4\pi\omega}{n_b c} \frac{1}{\pi a_H^2} \mathrm{Im} \left[ \sum_{\nu} \mu_{\nu,s}^\ast \chi_{\nu,s} \right] ~.
\end{eqnarray}
%%%
An example of the calculated gain for $B = 17$~T is shown in Figs.~5c and 5d.

%\bibliography{jun}

\end{document}